\documentstyle[12pt]{article}

\newcommand{\be}{\begin{equation}}
\newcommand{\ee}{\end{equation}}
\makeatletter
\@addtoreset{equation}{section}
\makeatother

\newcommand{\complex}{{\kern .1em {\raise .47ex
\hbox {$\scriptscriptstyle |$}}
    \kern -.4em {\rm C}}}
\newcommand{\real}{{{\rm I} \kern -.19em {\rm R}}}
\newcommand{\rational}{{\kern .1em {\raise .47ex
\hbox{$\scripscriptstyle |$}}
    \kern -.35em {\rm Q}}}
\renewcommand{\natural}{{\vrule height 1.6ex width
.05em depth 0ex \kern -.35em {\rm N}}}
\def\bea{\begin{eqnarray}}
\def\eea{\end{eqnarray}}
\def\unita{{1 \kern-.30em 1}}
\font\mybb=msbm10 at 12pt
\def\bb#1{\hbox{\mybb#1}}
\def\zet{\bb{Z}}
\def\cw{{\cal W}}
\def\cf{{\cal F}}
\def\de{\partial}
\def\Fey{\big / \kern-.96em {\ \cal{D}} }
\begin{document}
\begin{titlepage}
\begin{flushright}
{ROM2F/97/36}\\
\end{flushright}
\vskip 1mm
\begin{center} 
{\large \bf A Non-Technical Introduction to Confinement and
$N=2$ Globally Supersymmetric Yang-Mills Gauge Theories}\\   
\vspace{2.0cm}
{\bf F. Fucito}\\
{\sl I.N.F.N. Sezione di Roma II, 
Via Della Ricerca Scientifica, 00133 Roma, ITALY}
\vskip 3.0cm
{\large \bf Abstract}\\
\end{center}
The aim of this talk is to give a brief
introduction to the problem of confinement in QCD and to $N=2$
globally supersymmetric Yang-Mills gauge theories (SYM). While 
avoiding technicalities as much as possible I will try to emphasize
the physical ideas which lie behind the picture of confinement
as a consequence of the vacua of QCD to be a dual superconductor.
Finally I review the implementation of this picture in the 
framework of $N=2$ SYM.  
\vfill
\end{titlepage}
\addtolength{\baselineskip}{0.3\baselineskip}

\setcounter{section}{0}
\section{Introduction}
The last two decades have witnessed the successes of the
perturbative approach to the study of QCD. The discovery of
asymptotic freedom has marked an era of fruitful theoretical work
which has been throughly confirmed by experimental data.
Unfortunately one of the oldest experimental evidence in QCD,
the confinement of quarks and gluons inside hadrons, has not yet
received a satisfactory theoretical explanation. The main reason
for this fact is that confinement is a non-perturbative
effect which does not allow the use of a "standard" computational
technique. In spite of these difficulties great progresses have
been done in understanding non-perturbative phenomena in the past
years. I think that recalling few facts under a "historical"
perspective will help the reader follow my reasoning.
The first big progress has come with the discovery of
two non-trivial classical solutions of the equations of
motion of the Georgi-Glashow model and of pure gauge non-abelian
Yang-Mills theory: they
are called magnetic monopole and instanton respectively and will 
play a
major role in my future discussion \cite{col}. A magnetic monopole
is a field configuration of finite energy (and infinite
action) representing a particle with zero electric charge and
magnetic charge different from zero. As we will see later on, its
presence is badly needed in the description I will give of the
vacuum of QCD. An instanton is a field configuration of finite
action. Its importance rests on the fact that it allows
the evaluation of the contribution of tunnelling among different
vacua: loosely speaking instantons are the generalization to the
path integral formalism of the WKB method in quantum mechanics.
Soon after the instanton solution discovery a first attempt to evaluate
non-perturbative effects in QCD was carried out \cite{thoof1}.
In spite of the great technical difficulties overcome in this
work, the final result is disappointing, since it is infrared
divergent.
Some time earlier that this work it had appeared the proposal
of confinement as a dual Meissner effect \cite{thoof2} which
I will review in the next chapter. Even though this mechanism
is extremely appealing, its actual implementation is problematic
since the evaluation of the QCD effective action is a formidable 
task. The interest in the work that I will review in the third 
chapter rests on it being the first example of a non-trivial four
dimensional gauge theory in which the computation of the 
effective potential can be successfully carried out.
Let me now conclude this introduction by telling you the last part
of what I have called before a "historical" path. The last development
of interest for our story came afterwards the introduction
of  supersymmetry (SUSY). It was in fact very soon realized
\cite{vy} that SUSY greatly simplifies the evaluation of effective
actions as it is a very constraining symmetry. In its
presence, miracolous cancellations \cite{dv} avoid
the divergences found in \cite{thoof1}. Indeed SUSY is so
powerful that the final result of these computations can be
guessed in advance, by using Ward Identities, to be
a pure number times its physical dimension as it is later verified
by explicit computations \cite{asa}. Other important novelties
will come out in the presence of SUSY as we will see later.
But now our "historical" diversion is over and it is time to start
telling our story.

\section{Confinement as a Dual Meissner effect}
 
For a generic conductor, the current, $\vec j$, that flows into it, is
proportional to the electric field, $\vec E$ in which it is
immersed. The proportionality constant is called the
resistivity, $\sigma$
\be
\vec j=\sigma\vec E.
\label{uno}
\ee
In a superconductor $\sigma\to\infty$ and, to keep $\vec j$
constant, $\vec E\to 0$. As a consequence
\be
{\de\vec B\over\de t}=\vec\nabla\times\vec E\to 0.
\label{due}
\ee
Thus a superconductor expels
the magnetic field in which it is immersed. This 
is the Meissner effect and such a superconductor is called of type I.
If the external magnetic field is very strong, it
will penetrate into the superconductor (which will now be called
of type II), but
it will be squeezed into narrow flux tubes. If we imagine
the QCD vacuum to behave like a superconductor of type II, at the end
of the flux tubes we will have magnetic monopoles. It is well-known
that the phenomenon that sets the onset of the superconducting phase
is the creation of Cooper's pairs. For a field theorist the
way to describe this is via the Higgs mechanism: the inverse of the
mass thus generated, will be the size of the flux tube. What I have
described until now is the creation of magnetic flux tubes: color,
though, is an electric type of charge. 
The only mechanism at our disposal to exchange electric and magnetic
charges has to be of the type of the dual symmetry observed in abelian
theories. Maxwell's equation of motion are in fact invariant under
the transformation
\bea
&&\vec E\to \vec B,\nonumber\\
&&\vec B\to -\vec E.
\label{tre}
\eea
Then 
color confinement can be achieved only if in the QCD vacuum 
we see a kind of dual Meissner effect.
Appealing to duality is the only way out even if this is going to
raise a host of problems, since non-abelian
gauge theories are not invariant under a duality transformation.

As we have seen, the condensation of a pair of electrically charged
particles (via the Higgs effect), leads to the creation of a magnetic
flux tube. The dual of this effect, will require a magnetic Higgs
effect implying that the microscopic fields entering the Lagrangian
have to be magnetically charged: the description of the vacuum of QCD
thus imposes the existence of at least two different phases. How can
I characterize these two phases? The electric phase is usually
studied by introducing the Wilson line
\be
W(C^\prime)=Pe^{ig\oint_{C^\prime}A_\mu dx^\mu},
\label{quattro}
\ee
which "measures" the electric flux going through the surface
bounded by $C^\prime$. In the above formula, $g$ is the coupling
constant and $A_\mu$ is the non-abelian gauge connection of the
theory. For the magnetic phase we can think of introducing
an analogous order parameter, $M(C)$, "measuring" the magnetic
flux going through $C$. The main feature of the operator
$M(C)$ is to produce a singular gauge transformation
\footnote{That is to introduce a unity of magnetic flux or analogously
a monopole.}, $\Omega(C)$, 
upon acting on the vacuum of the theory, $|A_\mu>$
\be
M(C)|A_\mu>=|A^{\Omega(C)}_\mu>.
\label{cinque}
\ee
An explicit expression for $M(C)$ does
not exist in the framework of the second order formalism that consists
in writing the functional integral in terms of the field
strength $F_{\mu\nu}$ and with a functional measure given in
terms of $A_\mu$.
Upon introducing an auxiliary antisymmetric tensor field (first order
formalism) it is possible to write a non-local expression for $M(C)$
\cite{fmz}; in reality, the property of (\ref{cinque}) is sufficient
for the purpose I have in mind (to characterize the phases of QCD) so
I will not delve into this subject any longer. Before undertaking
our next computation, I remind the reader of the fact that a field
transforming in the adjoint representation of a gauge group (let's
say $SU(N)$) is blind to the center of the group so that
\be
\Omega(C):\qquad \Omega(2\pi)=e^{2\pi i{n\over N}}\Omega(0).
\label{sei}
\ee
I now compute
\bea
&&M(C)W(C^\prime)|A_\mu>=M(C)Pe^{ig\oint_{C^\prime}A_\mu dx^\mu}
|A_\mu>\nonumber=\\
&&Tr\{\Omega(2\pi)Pe^{ig\oint_{C^\prime}A_\mu dx^\mu}
\Omega^{-1}(0)|A^{\Omega(C)}_\mu>\}=e^{2\pi i{n\over N}}
Pe^{ig\oint_{C^\prime}A_\mu dx^\mu}|A^{\Omega(C)}_\mu>,
\label{sette}
\eea
and
\be
W(C^\prime)M(C)|A_\mu>=
Pe^{ig\oint_{C^\prime}A_\mu dx^\mu}|A^{\Omega(C)}_\mu>.
\label{otto}
\ee
Putting (\ref{sette}) and (\ref{otto}) together, I finally get
\be
W(C^\prime)M(C)=e^{ig\Phi}M(C)W(C^\prime),
\label{nove}
\ee
where $\Phi=2\pi n/N$. (\ref{nove}) represents a kind of braiding
relation taking in account the linkage between $C$ and $C^\prime$.
Up to now all is well, since I have worked in the Hamitonian
formalism at constant time. But going to a four dimensional Lagrangian
description I get into troubles since the linking of two curves
is well defined only in three dimensions. Using the fourth dimension
the two curves can be unlinked. The way out of this \cite{thoof3} 
is to replace the linkage between the two curves with the linkage
between one of the two curves and the surface, $\Sigma$, bounded 
by the
other curve, being this linkage well defined in four dimension.
The consequence is that the surface acquires a physical meaning and that
the effective action gets proportional to it. This is the origin
of the well-known area and perimeter laws in QCD. The options at our
disposal for the phases of QCD are
\begin{itemize}
\item
Higgs phase
$$
<W(C^\prime)>\simeq e^{-L(C^\prime)}\qquad;\qquad <M(C)>\simeq
e^{-\Sigma(C)}.
$$
\item
Confinement phase
$$
<W(C^\prime)>\simeq e^{-\Sigma(C^\prime)}\qquad;\qquad <M(C)>\simeq
e^{-L(C)}.
$$
\item
Partial Higgs phase
$$
<W(C)>\simeq <M(C)>\simeq e^{-\Sigma(C)}.
$$
\item
Coulomb phase
$$
<W(C)>\simeq <M(C)>\simeq e^{-L(C)}.
$$
\end{itemize}

\setcounter{equation}{0}
\section{$N=2$ Four Dimensional SYM}

In this section I will briefly review some recent work
on $N=2$ SYM \cite{sw} trying not to burden my account with
too many details which can be found in one of the many excellent
reviews that have appeared so far in literature \cite{rev}.
The starting point is the Lagrangian density of $N=2$ SUSY. In terms
of $N=1$ superfields, the $N=2$ multiplet contains
a vector and a chiral multiplet
\be
V=(\lambda^a_{\alpha 1},A_\mu^a,D^a)\qquad ;\qquad
\Phi=(\phi^a,\lambda^a_{\alpha 2},F^a).
\label{dieci}
\ee
The Lagrangian density of $N=2$ SYM (I choose $SU(2)$ as the gauge
group out of simplicity) is the same
of that of the $N=1$ theory coupled to matter in the adjoint
representation of the gauge group
\bea
{\cal L}&=&2Tr\bigg\{{1\over 4}F_{\mu\nu}F^{\mu\nu}+i\bar\lambda_A
\Fey\lambda^A+({\cal D}_\mu\phi)^\dagger({\cal D}^\mu\phi)\nonumber\\
&+&g^2[\phi,\phi^\dagger]^2+\bigg({ig\over\sqrt{2}}[\phi^\dagger
,\lambda_A]\lambda_B\epsilon^{AB}+h.c.\bigg)\bigg\}.
\label{undici}
\eea
The terms in the Lagrangian can be labelled as kinetic terms (those
containing covariant derivatives), couplings of Yukawa type (those
trilinear in the fields) and those giving rise to the potential
\be
V_{pot}=g^2[\phi,\phi^\dagger]^2.
\label{dodici}
\ee
The potential $V_{pot}$ has a very peculiar property: it has flat
directions that is a continuum of points for which the potential
attains its minimum, which has to be zero for the theory to be
SUSY. This is a characteristic of SUSY theories at large. In field
theory minima are points in general like those in the two wells
of a Higgs like potential. In (\ref{dodici}) every gauge rotated
scalar field will give a minimum: the minima of the theory are
a manifold and not isolated points. The condition $V_{pot}=0$
is satisfied by a normal field, which is conveniently chosen to be
\be
\phi^b=a\delta^{b3},
\label{tredici}
\ee
where $b$ is a gauge index.
Different values of the expectation value $a$, will lead to different
values for the masses of the gauge bosons and hence to theories with
different vacua. The space of these different vacua is called
the classical moduli space. One of our tasks will be of giving a
full description of the quantum moduli space, that is, to compute
the quantum potential of the theory. A manifold is conveniently
described upon the introduction of an appropriate set of coordinates.
It is very
natural to take as a coordinate of the classical moduli space
\be
u={1\over 2}Tr \phi^2={1\over 2}a^2,
\label{quattordici}
\ee
the simplest gauge invariant combination of the expectation value.
The quantum version of the moduli space will obviously use as a
coordinate the quantum expectation value of $Tr \phi^2$.

I need now some more $N=2$ technicalities: in first place let us
write down a $N=2$ superfield, $\Psi$, in terms of $N=1$ superfields
$\Phi, {\cal W}, {\cal G}$
\be
\Psi=\Phi(\tilde y,\theta)+\sqrt{2}+\tilde\theta {\cal W}(
\tilde y,\theta)+{1\over 2}\tilde\theta^2{\cal G}(\tilde y,\theta),
\label{quindici}
\ee
where
$y=x+i\tilde\theta\sigma\tilde\theta+i\theta\sigma\theta$.
A superfield is a convenient way
to bookeep the components of a SUSY multiplet: each member of the
multiplet will be a component of the superfield. As
in the multiplet  there will appear fields of different naive
dimensions (fermions and bosons) to write down an expansion of the
superfield\footnote{Remember that all the terms of 
the expansion must have the same naive dimension.} I introduce
a fermionic basis for each supersymmetry. The advantage of this way
of thinking is that a supersymmetric transformation is a translation
in this new space (called superspace) in which there are bosonic
and fermionic (as many as supersymmetries) directions
($\theta,\tilde\theta$) in (\ref{quindici}). 
A supersymmetric Lagrangian
is now a Lagrangian which is translationally invariant in the
superspace. The most generic SUSY $N=2$ Lagrangian is
\be
L={1\over 16\pi}\Im\int d^2\theta d^2\tilde\theta\cf(\Psi).
\label{sedici}
\ee
(\ref{undici}) is now recovered setting $\cf={\tau\over2}
\Psi^a\Psi^a$ as it can be checked by an explicit computation.
Expanding $\Psi$ around the $N=1$ superfield $\Phi$,
(\ref{sedici}) becomes
\be
L={1\over 16\pi}\Im\bigg[\int d^2\theta {\de^2\cf\over
\de\Phi^a\de\Phi^b}\cw^a\cw^b+
\int d^2\theta d^2\bar\theta(e^{2gV}\Phi^\dagger e^{-2gV})^a
{\de\cf\over\de\Phi^a}\bigg],
\label{diciasette}
\ee
where $V$ is the $N=1$ vector field.
The vacuum expectation value (\ref{tredici}) gives, via the Higgs mechanism,
a mass to the $W^\pm$ bosons of the theory, while the $Z^0$ stays massless
because the scalar is in the adjoint representation of the gauge group. It
is then possible to decouple the massive and massless particles of the theory
to study the effective Lagrangian
\be
L={1\over 16\pi}\Im\bigg[\int d^2\theta {\de^2\cf\over
\de\Phi\de\Phi}\cw\cw+
\int d^2\theta d^2\bar\theta\Phi^\dagger
{\de\cf\over\de\Phi}\bigg],
\label{diciotto}
\ee
which represents the contribution of the massless states only. As our gauge group
was $SU(2)$ the only massless degree of freedom has a $U(1)$ symmetry.

Before proceeding further I now take a little time to explain the physical meaning
of the effective action (\ref{diciotto}). 
Take, as an example, QED with a cut-off $\Lambda_0$. The Lagrangian is
\be
L_0=\bar\psi(i\de\cdot\gamma-e_0A\cdot\gamma-m_0)\psi-{1\over 2}(F_{\mu\nu})^2.
\label{diciannove}
\ee
The idea is that if I use a new cut-off $\Lambda\ll\Lambda_0$ the new Lagrangian
can be obtained from that of (\ref{diciannove}) by adding a certain number of terms
which represent new interactions \cite{lepage}. For example, from the one loop diagram,
arising from the scattering of two fermions, of momenta $p, p^\prime\ll\Lambda$,
off the external field, I get
\bea
T&=&-e_0^3\int_{\Lambda}^{\Lambda_0}{d^4k\over(2\pi)^4}{1\over k^2}\bar u(p^\prime)\gamma^\mu
{1\over(p^\prime-k)\cdot\gamma-m_0}A(p^\prime-p)\cdot\gamma{1\over(p-k)\cdot\gamma-m_0}
\gamma_\mu u(p)\nonumber\\
&\simeq& -{ie_0^3\over 24\pi^2}\ln({\Lambda\over\Lambda_0})\bar u(p^\prime)
A(p^\prime-p)\cdot\gamma u(p).
\label{venti}
\eea
This contribution can be taken care of by adding to the Lagrangian, the term
\be
\delta L=-{e_0^3\over 24\pi^2}\lg({\Lambda\over\Lambda_0})\bar\psi A\cdot\gamma\psi,
\label{ventuno}
\ee
which amounts to a renormalization of the electric charge. Other terms of higher order in
$p/\Lambda$ arising from (\ref{venti}) and coming from other diagrams may be added to the
effective Lagrangian (in fact there is an infinite number of such terms)\footnote{All the 
terms I add according to this procedure have the same symmetries of the Lagrangian 
(Lorentz, gauge etc.) and are local. The non-locality of the effective Lagrangian is now
given by the terms of order $(p/\Lambda)^n$ with any $n$.} 
Some of them 
will be non  renormalizable, given their naive dimensions. In this approach this does not raise
any concern. {\it The Lagrangian represents the theory at a certain energy scale given by the
cut-off}. This is the same thing of what happens in strong interactions when, to describe the 
theory, I use a $\sigma$-model, where the fundamental fields are the mesons, instead of using
the standard QCD Lagrangian. Coming back to our problem, (\ref{undici}) is the analogous
of (\ref{diciannove}) that I have written as (\ref{diciasette}) to allow for the next step.
Going to an energy scale (cut-off) much lower than the masses of the
particles corresponds to use the effective Lagrangian (\ref{diciotto}). The task is now to
explicitly build it. Trying to do it by computing the relevant diagrams as in the previous example,
is hopeless. Luckily enough the symmetries of the theory and an educated ansatz will be enough to do
the job as we will see later on.

I now want to convince the reader that the functional $\cf(\Phi)$ is in reality
a multi-valued function so that, to build it, all I have to do is to know its monodromies
around its singular points. In terms of its component fields, (\ref{diciotto}) is
\be
L={1\over 4\pi}\Im\bigg[\cf^{\prime\prime}|\de_\mu\phi|^2-i\cf^{\prime\prime}
\psi\sigma^\mu\de_\mu
\bar\psi-{1\over 4}\cf^{\prime\prime}F_{\mu\nu}F^{\mu\nu}+\ldots\bigg],
\label{ventidue}
\ee
where I have written only the terms which will be necessary for our discussion.
From (\ref{ventidue}) I see that $\tau=\cf^{\prime\prime}$ behaves as the metric
of a $\sigma$-model and that it has to be positive as it multiplies the kinetic terms.
Then $\tau$ is a holomorphic function which must be positive. From a well-known theorem
in complex analysis it then has to be a constant. This happens in the classical case,
but I have seen that in the effective action, (\ref{diciotto}), it is not true anymore:
therefore the coupling $\tau$ which appears in (\ref{diciotto}) must be a multi-valued function.
The monodromies of this function will be dictated by the type of physics 
to be described.

We have seen in the second chapter that the picture of confinement as a dual Meissner effect,
requires at least an electric and a magnetic phase which are connected by a duality transformation.
Let us build now this transformation in our case. The commutation relations of the SUSY
charges of the $N=2$ algebra lead to the remarkable formula \cite{ow}
\be
M=a|q+ig|,
\label{ventitre}
\ee
where $M$ is the mass of the particle and $q, g$ its electric\footnote{Electric here as to be
interpreted as the coupling of the theory appearing in the Lagrangian describing the
electric phase.}
and magnetic charge. As this
relation comes out of the algebra, it is valid both at the classical and quantum level.
Moreover we see that using the Dirac quantization rule, $qg=4\pi$, (\ref{ventitre}) is left
invariant. In reality the dual transformation I am after is more involved that the Dirac rule.
It is in fact possible to show \cite{witten2} that the appropriate electric coupling of 
the theory is
\be
\tau=\bigg({4\pi i\over q_0}+{\theta\over 2\pi}\bigg),
\label{ventiquattro}
\ee
where $\theta$ is the coupling I have to add to the theory to take into account the effect of
instantons. If I define $q=nq_0, g=mg_0$, where $q_0, g_0$ are the fundamental units of charge,
and redefine $a=aq_0, a_D=\tau a$, then (\ref{ventitre}) becomes
\be
M=|an+a_Dm|.
\label{venticinque}
\ee 
(\ref{venticinque}) describes a lattice on whose sites are the charges. The symmetry
group of this lattice is $SL(2,\zet)$, whose most general transformation is
\be
\tau^\prime={a\tau+b\over c\tau+d},
\label{ventisei}
\ee
with $a,b,c,d\in\zet$.
The generators of $SL(2,\zet)$ are
\bea
T&:&\qquad \tau\mapsto\tau+1\nonumber\\
S&:&\qquad\tau\mapsto -{1\over\tau}.
\label{ventisette}
\eea
The transformation $S$ is the analogous of the dual transformation given by the Dirac rule
in which strong and weak coupling are exchanged. In order to implement these transformations
in our Lagrangian (\ref{diciotto}), I need the quantum expression for $a_D$.
If classically $\cf=1/2\tau a^2$ then
\be
a_D=\tau a={\de\cf\over\de a},
\label{ventotto}
\ee
which I take to be valid in the quantum domain once $\cf$ is substituted by its quantum
expression. If I extend the definition (\ref{ventotto}) to the entire superfield of which
$a$ is the scalar component, I get $\Phi_D=\cf^\prime(\Phi)$.
At the level of the Lagrangian, the transformations $S$ is given by
\be
\pmatrix{\Phi_D\cr\Phi}\mapsto\pmatrix{0&1\cr -1&0}\pmatrix{\Phi_D\cr\Phi}=
\pmatrix{\Phi\cr\Phi_D}.
\label{ventinove}
\ee
Transforming the Lagrangian according to (\ref{ventinove}) and using the Legendre transform
\be
\cf_D(\Phi_D)=\cf(\Phi)-\Phi\Phi_D,
\label{trenta}
\ee
I recover a new Lagrangian which has the same functional form of the original one with
the electric fields exchanged with the magnetic ones and a new magnetic coupling
$\tau_D=-1/\tau$. The $T$ transformation is given by
\be
\pmatrix{\Phi_D\cr\Phi}\mapsto\pmatrix{1&b\cr 0&1}\pmatrix{\Phi_D+b\Phi\cr\Phi}=
\pmatrix{\Phi\cr\Phi_D},
\label{trentuno}
\ee
where $b\in\zet$, and it leaves the Lagrangian invariant.

I am now ready for the computation of the effective action. The strategy is very simple.
I give an ansatz for the moduli space to make it obey the type of physics I want to describe.
I then compute the monodromies of $\cf$ around the singularities of the moduli space. Knowing
these singularities I find $a, a_D$. So what is the minimum number of singularities I can have?
The moduli space is described by a complex variable, so it is two-dimensional. Then the simplest
Riemann surface I can build, on which $\cf$ is single-valued is the thrice punctured sphere or the
torus. One of the singularities will be given by the weakly coupled ( that is around $a=\infty$)
Higgs phase; in this case I can exhibit the form of $\cf$. One other phase will be the weakly
coupled  magnetic phase (around $a_D\simeq 0$). The theory around this point has to look like
a magnetic QED, that is as a $U(1)$ theory in which the fundamental fields are the monopole
and a dual photon. This is implicit in the idea of duality for which to create a magnetic 
condensate I need a magnetic Higgs mechanism. It is remarkable that in this description
the monopole naturally comes to have the right mass for the Higgs mechanism, without having
to introduce special gauges \cite{thooft4}.
In the weakly coupled electric phase $\cf$ looks like
\be
\cf={i\over 2\pi}a^2\ln{a^2\over\Lambda^2}+
\sum^\infty_{k=1}\cf_k\bigg({\Lambda\over a}\bigg)^{4k}a^2.
\label{trentadue}
\ee
The first term in the r.h.s. is the perturbative contribution due to the $U(1)$ anomaly,
while the sum is given by the instanton contribution. In the $a\to\infty$ limit, the
perturbative sector dominates the non-perturbative one. Then if $u\to exp\{2\pi i\}u$,
the monodromy is given by
\be
\pmatrix{a_D\cr a}=\pmatrix{-1&2\cr 0&-1}\pmatrix{a_D\cr a}=M_\infty\pmatrix{a_D\cr a}=
\pmatrix{-a_D+2a\cr -a}.
\label{trentatre}
\ee
In the $U(1)$ magnetic phase I will have that the $\beta$ function of the theory is
\be
\beta(g)=-{g^3\over 4\pi^2},
\label{trentaquattro}
\ee
which goes to zero for $g\to 0$.  The monodromy around this point is then
\be
\pmatrix{a_D\cr a}=\pmatrix{1&0\cr -2&1}\pmatrix{a_D\cr a}=M_1\pmatrix{a_D\cr a}=
\pmatrix{a_D\cr a-2a_D}.
\label{trentacinque}
\ee
The monodromy around the third singularity (that I postulate to be a dyon with
one unit of electric and magnetic charge) is given by the group relation
\be
M_\infty=M_1M_{-1}.
\label{trentasei}
\ee
Finding the $a, a_D$ with these monodromy is now an exercise in complex analysis,
whose solution is
\bea
a&=&{\sqrt{2}\over\pi}\int_{-1}^1{\sqrt{x-u}\over\sqrt{x^2-1}}dx,\nonumber\\
a_D&=&{\sqrt{2}\over\pi}\int_{1}^u{\sqrt{x-u}\over\sqrt{x^2-1}}dx.
\label{trentasette}
\eea

Before concluding this chapter, I want to show that in the magnetic phase, the monopole field
develops a vacuum expectation value. As I have already said, 
in this phase the Lagrangian is given by a QED type Lagrangian
\bea
L&=&{1\over 16\pi}\Im\bigg[\int d^2\theta {\de^2\cf_D\over
\de\Phi_D\de\Phi_D}\cw_D\cw_D+
\int d^2\theta d^2\bar\theta\Phi_D^\dagger
{\de\cf_D\over\de\Phi_D}\bigg]\nonumber\\
&+&\int d^2\theta d^2\bar\theta[\tilde Q^\dagger e^{-2gV_D}\tilde Q
+Q^\dagger e^{-2gV_D} Q]\nonumber\\
&+&\int d^2\theta [\sqrt{2}\Phi_DQ\tilde Q+mU(\Phi)].
\label{trentotto}
\eea
All the fields with the subscript $D$ are the dual of those appearing in (\ref{diciotto}).
The second line contains the monopole superfields $Q=(q,\psi_Q,F_Q) 
\tilde Q=(\tilde q,\tilde\psi_Q,\tilde F_Q)$ with the typical matter 
interaction\footnote{I remind the reader who is not expert in SUSY, that there are two Weyl
fields in the description of SUSY QED to reproduce the Dirac spinor appearing in the standard
QED Lagrangian.}. At last,  the third line contains the only possible form of an interaction
which is compatible with $N=2$ SUSY \cite{sohnius} and a mass term which breaks $N=2$ into
$N=1$. As we will see shortly, this term is necessary to develop a vacuum expectation value.
As (\ref{trentotto}) is an effective Lagrangian, according to our previous discussion,
multiplying the mass term there is a generic non-renormalizable function
$U(\Phi)$ instead of the standard $\Phi^2$ term.
The potential of this theory in components is
\bea
V&=&\sqrt{2}[F_Dq\tilde q+a_DF_Q\tilde q+a_DqF_{\tilde Q}]+m U^\prime(a_D)F_D+h.c.\nonumber\\
&+&F_Q\bar F_Q+F_{\tilde Q}\bar F_{\tilde Q}+|q|^2D+|\tilde q|^2D+\Im\tau_D\bar F_DF_D+
{1\over 2}D^2.
\label{n39}
\eea
To find the minimum of the potential I just minimize with respect to the auxiliary fields.
As the potential has to be zero to conserve SUSY I see that $F_Q=F_{\tilde Q}=F_D=D=0$.
Plugging this value back into the minimization with respect to the auxiliary fields, I find the
minimum of the potential for
\be
q=\bigg({m\over\sqrt{2}}{dU(a_D)\over da_D}\bigg)^2\vert_{a_D=0}\neq 0,
\label{n40}
\ee which is the sought for result. I point out that the expectation value for the monopole,
goes to zero for $m\to 0$, as I remarked before.

\section{Conclusions and Overlook}
The work I have just described, has triggered an intensive activity 
in the last years. Before attempting a rough description of these developments, 
I would like to comment on the physical relevance of the results I have discussed: this is
the first example of a four dimensional gauge theory in which an almost complete analysis
of the perturbative and non-perturbative sector is carried out. In the past similar analysis
were performed on $O(N)$ or $\zet_N$ Ising type models, whose similarity with QCD was much less evident
than for SUSY YM models. In this lecture, all the ingredients that seem to be important for
confinement are at work: duality, massless monopoles, phases of QCD etc. Moreover, for the
first time, an extensive control of the non-perturbative sector has been achieved. All of this
has been possible because of SUSY and of the peculiar place that $N=2$  has among
all SUSY models. The $N=1$ model has in fact a perturbative sector which is by far more complicated
than the one discussed here. At the same time $N=4$ has a trivial non-perturbative sector.
$N=2$ seems to have that right mixture that makes it rich enough to show the interesting
phenomena but not too constrained to become trivial. On the relevance of SUSY for physics I
won't comment at all for lack of time. I just want to say that if hints of SUSY are not going
to be found in the next experiments, most of the work done in the last twenty years is in peril,
since SUSY is now obiquitous in most of theoretical physics.

The developments of the ideas exposed in this lecture, may be divided into two main streams:
studies of duality and applications to string theory, on which I will not comment, and 
further studies in global $N=2$. In first place the results presented here have been
extended to generic gauge groups \cite{klt}.
Then the generality of the ansatzes used for the solution presented \cite{m1} here and its relation
with microscopic instanton calculus \cite{ft} have been checked. It is remarkable that the quantum
symmetry of the theory is sufficient to make the solution I have discussed here the only possible 
one and to give results which are in agreement with instanton computations. Also the
non-holormorphic sector of the theory has been recently investigated and the preliminary
results seem to be encouraging \cite{bfmt}. Finally there have also been attempts to extend
the results to $N=0$ \cite{am}.
\leftline{\large\bf Acknowledgments}
The author wants to thank N.Berkovitz and C.Lozano for pointing out to him
some misprints and an error in (\ref{diciasette}).

\newpage

\end{document}